\documentstyle[twocolumn,aps,prl]{revtex}

\begin{document}
%\draft
\def\beq{\begin{equation}}
\def\eeq{\end{equation}}
\def\beqn{\beqin{eqnarray}}
\def\eeqn{\end{eqnarray}}
\newcommand{\bkbox}{(Ba$_{1-x}$K$_{x}$)BiO$_{3}$}
\newcommand{\etal}{{\em et al.}}
\newcommand{\Tc}{T$_{c}$}

\title{A Comment on ``Superconducting-Normal Phase Transition in  
(Ba$_{1-x}$K$_{x}$)BiO$_3$, x = 0.40, 0.47'' by B. F. Woodfield, D. A. 
Wright, R. A. Fisher, N. E. Phillips and H. Y. Tang, Phys. Rev. Lett. 
\bf{83}, 4622 (1999)}
\author{P. Kumar} 
\address{Department of Physics, University of Florida, Gainesville,
FL  32611}
\author{Donavan Hall} 
\address {National High Magnetic Field Laboratory, 
Florida State University, Tallahassee, FL 32310}
\author{R.~G. Goodrich} 
\address {Department of Physics and Astronomy, Louisiana State
University, Baton Rouge, LA  70803--4001}

\date{\today}
\maketitle

%\narrowtext
%\twocolumn

In a recent paper by Woodfield \etal \cite{Woodfield1999}, comments
have been made on our earlier paper \cite{Kumar1999}, their reference
[1].  In this paper it is stated that: (a)  that there is specific
heat discontinuity at the superconducting phase transition in \bkbox~
(for x = 0.4 and 0.47) of the order of a few mJ/mole-K, (b)  that this
is what is expected, and (c)  that there is no reason to invoke a
higher order transition, as we recently suggested.

The logical foundation of our paper, which suggests that the
superconducting-normal transition in \bkbox~ with x = 0.40 may be of order
IV, rests on three independent observations.  They are: (1) the lack
of an observed discontinuity in magnetic susceptibility at the
superconducting transition (\Tc), $\Delta \chi$ = 0, (2) near \Tc, the
thermodynamic critical field, H$_{0}$(T), obtained by the integration
of the magnetization M(H) at different temperatures, depends on
temperature as (1-T/\Tc)$^{2-\mu}$, where $\mu$ is small and $<$ 1
and, (3) the lower critical field near \Tc ~fits the expression
H$_{c1}$(T) $\propto$ (1-T/\Tc)$^{3}$.

The fact that, in all of the samples
we have measured, the M vs.  H curves (1) never approach 
H$_{c2}$ linearly,
as required for a second order transition from the Abrikosov state 
\cite{Tinkham},
and (2) the slope varies smoothly into the normal state, substantiates 
our first observation stated above.  Contary to the
statement by Woodfield \etal, that a discontinuity was observed by
Hundley \etal \cite{Hundley1989}, in their constant field temperature dependent
suseptibility measurements, they did see a discontinuity at T$_{c1}$ where
complete flux exclusion occurs but no discontinuity at T$_{c2}$, the normal
to superconducting transition.  Thus, observation (1) is consistent
with a vanishing specific heat discontinuity, $\Delta$C = 0 and implies that
the transition \emph{cannot be second order}.  The assertion for
a IV order transition comes from observation (2).  In the end,
observation (3) is a verification of a model for a IV order phase
transition.  \emph{All three of the experimental observations are
inconsistent with a second order phase transition.}  If the transition
were II order, then the exponents in (2) and (3) would be,
respectively 1-$\alpha$, and 1 where $\alpha$ is the small specific heat exponent.
  
In addition, there had not been a
finite value of $\Delta$C at \Tc~measured in the published data of
Hundley \etal \cite{Hundley1989} and Stupp \etal \cite{Stupp1989} 
even though their results indicate that, in the
normal state $\gamma$ = 150 mJ/moleK$^{2}$.  This large value of 
$\gamma$ made the problem look acute in that the difference between 
the expected size of the discontinuity and the experimental 
uncertainty was large. In Woodfield \etal, there is no mention of
the volume fraction of the samples that become superconducting, making
the magnititude of their estimates of $\gamma \sim$ 1 mJK$^{-2}$ of unknown accruacy
even if the transition were second order.

The expectation that $\Delta$C $\sim$ 1 mJK$^{-2}$ is based on a BCS 
expression which is to assume that the transition is second order.  
For example, if we put the observed temperature dependence of 
H$_{0}$(T) in the expression used by Batlogg \etal \cite{Batlogg1988b}, we get $\Delta$C = 
0.  Similarly, the expressions described in Ref. 
\onlinecite{Hundley1989} are also suspect because they assume a value 
for $\kappa$ which in light of its temperature dependence leads to 
$\Delta$C = 0.

Interpreting
the data of Woodfield \etal~ is problematic because of apparent temperature 
independence of the anomaly.  For the highest \Tc, x = 0.4, samples the
magnitude of the specific heat anomaly changes with field, but the
temperature location of the anomaly changes very little.  There is a
large amount of scatter in the data presented, but one can argue that
the peak in C for H = 1 T is actually higher in temperature than the
one for H = 0.5 T. In addition, there appears to be no anomaly at any
temperature for fields $>$ 3 T even though H$_{c2}$ is only reduced in
all other measurements to about 22 K at 5 T, and exceeding 20 T at 4 K.
For the x = 0.47 sample, no peak is observed for fields $>$ 0.5 T, and
we point out that the temperature of the observation of the peaks is
in the range, 12 - 17 K, where we previously noted anomalies in the
critical field vs.  temperature curves.  \emph{We must conclude that
whatever this data may represent, it may not be the onset of
superconductivity.}

A portion of this work was perfomed at the National High Magnetic Field
Laboratory, which is supported by NSF Cooperative Agreement No. DMR-9527035
and by the State of Florida.

%\bibliographystyle{prsty}
%\bibliography{bkbophys}

\end{document}